\documentclass[manuscript]{aastex}

\begin{document}
\title{X-ray Emission from the Warm-Hot Intergalactic Medium}


\author{E. Ursino and M. Galeazzi\altaffilmark{1}}
\affil{Physics Department, University of Miami, Coral Gables, FL 33155}
\altaffiltext{1}{corresponding author, galeazzi@physics.miami.edu}


\begin{abstract}
The number of detected baryons in the Universe at $z<0.5$ is much smaller 
than predicted by standard  big bang nucleosynthesis and by the detailed 
observation of the Lyman alpha forest at red-shift $z=2$. 
Hydrodynamical simulations indicate that a large fraction of the baryons 
today is expected to be in a ``warm-hot'' ($10^5$-$10^7$~K) filamentary gas, 
distributed in the intergalactic medium. This gas, if it exists, should be 
observable only in the soft X-ray and UV bands.
Using the predictions of a particular hydrodynamic model, we simulated the 
expected X-ray flux as a function of energy in the 0.1-2~keV band due to 
the Warm-Hot Intergalactic Medium (WHIM), and compared it with the flux from 
local and high red-shift diffuse components. Our results show that as much 
as 20\% of the total diffuse X-ray background (DXB) in the energy range 
0.37-0.925~keV could be 
due to X-ray flux from the WHIM, 70\% of which comes from filaments 
at redshift z between 0.1 and 0.6. Simulations done using a FOV of 3', 
comparable with that of Suzaku and Constellation-X, show that in 
more than 20\% of the observations we expect the WHIM flux to 
contribute to more than 20\% of the DXB. These simulations also show that 
in about 10\% of all the observations a single bright filament in the FOV 
accounts, alone, for more than 20\% of the DXB flux. Red-shifted 
oxygen lines should be clearly visible in these observations. 
\end{abstract}



\section{Introduction}

The detailed observation of the Lyman alpha forest at red-shift $z=2$, the 
result from the WMAP experiment, and the observation of the light element 
ratio combined with standard nucleosynthesis, all point to the baryon 
density (expressed as a present day fraction of the closure density of 
the universe) $\Omega_B$, being about $ 0.04$ \cite{Rauch98, Weinberg97, 
BurTyt98, Kirkman03, Bennett03}. In contrast, integrating the observed
 mass distribution function 
of stars, galaxies and clusters, the baryonic fraction of the local universe 
 \cite{Fukugita98}($z\sim0$) appears to be about a factor of 2 to 4 times 
lower.

Where is the missing baryonic mass at $z\sim0$? The state of these missing
 baryons can be computed from standard initial conditions in realistic 
large-scale cosmological hydrodynamic simulations \cite{CenOst99, Croft01, 
Borgani04, Yoshikawa03}. Both theoretical and recent observational
 work \cite{Tripp02} suggest that much of the ``missing'' 
material lies in a ``warm-hot'' filamentary gas, distributed in the 
intergalactic medium (WHIM), with temperature between $10^5$ and $10^7$~K,
 which makes it very difficult to observe. The best theoretical estimates 
of its distribution in space and temperature indicate that the bulk of the 
material lies in regions with a density between 20-1000 times the 
average density of the universe, and has a temperature greater than 
$5\times10^5$~K, which essentially makes it invisible to all but low 
energy X-ray observations.

There have been several recent measurements that offer tantalizing hints 
on the existence of this material. Both Chandra \cite{Nicastro02, Fang02} 
and XMM \cite{Rasmussen02, Rasmussen03} grating observations have detected 
resonance O~{\tiny VII} and O~{\tiny VIII} absorption lines along the
 lines of sight 
to several distant quasars and BL~Lac objects. Most of the observed 
features lie near zero velocity, suggesting that they may be caused by hot 
gas in our own Galaxy or in the Local Group. Simple estimates of the mass 
and temperature 
of this material indicate that it may dominate the baryon content of the 
Local Group and has a temperature near $2\times10^6$~K, consistent with the 
virial temperature of the Local Group. Assuming that the Local Group 
is representative of other spiral-rich groups, which contain the bulk 
of all galaxies in the universe, these observations alone would more 
than double the census of observable baryons. As shown by Mulchaey et 
al. (2003), the existence of such low temperature flux would explain
 much of the ROSAT data for spiral-rich groups. ROSAT imaging and 
spectroscopy \cite{Zappacosta02, Scharf00} suggest the existence of 
several X-ray ``filaments'' or clumps that are not obviously associated
 with rich clusters or groups. Their properties are similar to those 
predicted in numerical simulations of the Intergalactic Medium (IGM), but 
the low quality of the 
data prevents strong conclusions. A recent observation \cite{Nicastro05} 
found absorption lines at redshift significantly greater than 0 
(z=0.011, z=0.027). This seems to indicate the presence of absorbers 
out of the Local Group. These data can be used to estimate the abundance 
of filaments on a line of sight as $dP/dz = 67_{-43}^{\phantom{-43}+88}$, 
however only one 
observation is available at this point and the sensitivity of current 
experiments does not permit to improve the statistics past this one data 
point. 

In addition to the resonant absorption line in the lines of sight of 
distant quasars and BL Lac objects just mentioned, three different 
approaches have been proposed for detecting the WHIM. All methods rely on 
the detection of the X-ray flux from the WHIM on top of the 
foreground due to the flux from the Local Bubble and the Galactic halo
 and the background due to very distant point sources.

\begin{enumerate}
\item	Red shifted ($z < 1$) strong soft X-ray emission lines from highly 
ionized elements (such as O~{\tiny VIII} 653~eV line) should be observable 
\cite{Galeazzi00, Galeazzi04}. 

\item	The filamentary structure of the WHIM should generate a 
characteristic angular pattern in the soft X-ray flux that can be detected 
using the angular autocorrelation function \cite{Soltan96, Soltan01, 
Croft01}.

\item	Search for shadows of the more distant WHIM cast by nearby 
galaxies that are rich in neutral hydrogen \cite{Barber96, Bregman01}. 
\end{enumerate}

The first of the three methods is the one that promises the best results 
since, in addition to the detection of the WHIM, it could provide also a
 tool to study the characteristics of the gas. In this work we used 
the prediction of the hydrodynamic model of Cen and Ostriker (1999, 
Cen99 from here on) to 
simulate the expected X-ray flux due to the WHIM, to determine its 
detectability and expected characteristics. This will offer a tool to 
determine the effect of the WHIM on future and present missions and to
 possibly design a future mission dedicated to the study of the WHIM. 
Our investigation has been focused on the soft X-ray 
spectrum between 50~eV and 2000~eV, with special attention to the O 
{\tiny VII} 
and O~{\tiny VIII} lines. The results are in the form of energy spectra and 
images. It has been possible to derive general characteristics of 
the WHIM, and how the WHIM may be detected and may affect future experiments.

\section{The Model}

The starting point of our simulations are the results of the hydrodynamic 
simulation of Cen and Ostriker (Cen99). 
Among other results, this simulation gives a set of three cubes 
100~Mpc~$h^{-1}$ 
in size [$h=H_0/(100 \textrm{~km~s}^{-1}\textrm{~Mpc}^{-1}$), with 
$H_0$ being the current time 
Hubble constant], with $512\times512\times512$ elements, containing 
temperature, metallicity (as fraction of the solar one), and 
overdensity ($\rho/\rho_0$, with $\rho_0$ being the critical density 
of the Universe) for
 the intergalactic medium at z=0 (Cen's cubes from here on). We used 
Cen's cubes to build a 3D representation of the intergalactic medium 
up to z=2 piling up several cubes. As will be discussed in more details 
later on, we didn't go to higher redshift because there is no significant 
soft X-ray flux beyond z=2. To limit the effect of periodicity 
(i.e., finding the same structure in the same position every time), each 
cube is randomly rotated, shifted, and the coordinates are permutated, 
so that each cube could have 24,576 different configurations. The values 
of temperature, metallicity and overdensity of each element are also 
rescaled depending on the red-shift of the element, following the expected 
average changes reported by Cen and Ostriker (1999, 2001).

To determine the X-ray flux due to the WHIM, an arbitrary starting point
is chosen in the first cube and a conical field of view (FOV) is generated
with an angular aperture that can be arbitrarily chosen at the beginning of the
simulation.
Among all the elements that fit in the field of view, only those belonging
to the WHIM are selected. For the selection we followed the indication
of Cen and Ostriker (Cen99), and excluded elements with temperature above
$10^7$~K and/or overdensity above 1000 (clusters of galaxies), and elements
with temperature below $10^5$~K (cool gas). This selection works
quite well at identifying clusters of galaxies and other regions that don't
contribute to the X-ray flux from the intergalactic medium. However, in our
analysis we observed a residual flux of X-rays coming from the outer
regions of galaxy clusters. Our results may therefore somewhat overestimate 
the true intergalactic flux.
For each element with values of temperature and density within our
acceptable parameters, an X-ray spectrum is generated. The spectra are then
added together to generate the X-ray spectrum due to the WHIM.
The spectra are generated using the APEC thermal model included in the code
XSPEC \footnote{http://heasarc.gsfc.nasa.gov/docs/xanadu/xspec/}.
Even for a FOV of a few arcmin, several million elements are included in
the FOV, making the call to XSPEC for each element unfeasible.
We therefore generated a matrix of 1470 spectra, using forty-nine different
values of temperature equally spaced on a logarithmic scale from
$2.5\times10^5$~K to $1.18\times10^8$~K, and thirty values of metallicity,
equally spaced on a logarithmic scale from $5\times10^{-5}$ to 5 (expressed 
as fraction of the solar metallicity),with the higher limit due to the APEC 
model limitations. The spectra were generated using solar abundances, according
to the model by Anders and Grevesse (1989).
Each spectrum is then obtained as the weighted average
of the four spectra in our matrix with values of temperature and
metallicity closer to the element values.
For elements at the edge of the FOV, the spectral amplitude is also scaled
based on the fraction of the element that fits in the FOV. This is
particularly important for small FOV's, and for elements at low redshift.
Failing to perform this rescaling, would result in a significant
overestimate of the WHIM X-ray flux.
Galactic absorption is applied at the end of the simulation using the
model by Morrison and McCammon (1983).
For the current analysis, we adopted a typical high latitude column
density value of $1.80\times10^{20}\textrm{~cm}^{-2}$ \cite{McCammon02}.

\section{Results}

For the analysis we adopted three different FOV's, equal to 
3', 10', and 30'. A FOV of 3' is comparable to what is expected for 
Constellation-X~\footnote{
http://constellation.gsfc.nasa.gov/mission/overview/index.html}, 
while 30' is comparable to that of XMM-Newton 
\footnote{http://xmm.vilspa.esa.es/external/xmm\_user\_support/documentation/uhb/node14.html}. 
The FOV of 10' has been chosen as an intermediate 
value.

Most of the results reported here focus on the energy interval 
$0.375 - 0.950$~keV. The upper limit of this energy range has been chosen 
to include  the emission from Ne~{\tiny IX} and the strong 
Fe~{\tiny XVII} lines from 0.725 to 0.827~keV. The lower 
limit has been chosen as to consistently exclude the strong 
C~{\tiny VI} lines at 0.367~keV. This makes us sensitive to 
O~{\tiny VII} out to redshifts of z=0.5.~A lower limit well 
above 0.284~keV is also advantageous to avoid instrument effects due 
to the neutral carbon absorption edge which is present in most instruments. 
Our simulations indicate that, in this energy band, the average flux due to the WHIM is $6.9\pm0.9 
\textrm{~photons~cm}^{-2}\textrm{~s}^{-1}\textrm{~sr}^{-1}$. 
Compared with the expected total flux in the same energy range due to the DXB of 29.3
$\textrm{~photons~cm}^{-2}\textrm{~s}^{-1}\textrm{~sr}^{-1}$ (derived using 
data from McCammon et al. 2002, Table 3), our result indicates that as much 
as 20\% of the diffuse X-ray background could be due to X-ray flux 
from the WHIM. This is an upper limit to the X-ray flux from the WHIM as,
with our data selection, it may still include flux from the outer regions of 
galaxy clusters. 

We also studied the evolution of the WHIM flux with increasing redshift. We 
split the column along the line of sight in 30 intervals, each interval 
corresponding to one Cen's cube (this is not a linear spacing in redshift) and
we saved the data of the flux calculated 
for each interval. Figure~1 shows the distribution of the WHIM flux as a 
function of redshift. We found that, in the energy range we used for our 
analysis, the X-ray flux due to the WHIM is approximately constant with 
redshift up to $z\sim 0.5$ and that 70\% of the total flux comes from gas
at redshift between 0.1 and 0.6. At redshifts higher than 0.6 the X-ray flux 
drops quickly both because emission lines are shifted outside the energy range 
and because the density of the gas becomes significantly smaller. In 
particular, the two steps in the X-ray flux at $z\sim 0.5$ and $z\sim 0.7$ 
are due to the fact that at those values the O~{\tiny VII} and O~{\tiny VIII} 
lines respectively are redshifted outside the energy range. The figure also 
shows that, in the energy range considered, there is no significant X-ray 
emission from gas at redshift higher than 1.

Of particular interest for the design of a mission to design the properties 
of the WHIM is how the X-ray flux changes for different FOV's.
While the average of the flux over several simulation obviously does not depend
on the FOV, the distribution of the flux from individual simulations strongly
depends on it. In Figure 2 we reported the flux distribution of the simulations
for the three FOV's. This gives the probability that an observation would 
detect a specific flux due to the WHIM.  In the case of 3' FOV most 
observations don't 
encounter any bright X-ray emitting region (a ``filament'' of the WHIM) 
and the X-ray flux is close to zero. When a bright filament is encountered, 
however, the flux can reach very high values. The distribution of the flux 
per run looks almost exponential, with a maximum for very small flux 
(around 1 $\textrm{~photon~cm}^{-2}\textrm{~s}^{-1}\textrm{~sr}^{-1}$), 
and with a reasonable number of simulations that show a very high X-ray 
flux. In the case of 30' FOV, instead, several filaments tend to always be 
in the FOV and the distribution is much narrower and peaked around the 
average value of $6.9\pm0.9\textrm{~photons~cm}^{-2}\textrm{~s}^{-1}
\textrm{~sr}^{-1}$. The distribution of the 10' FOV runs is in between 
these two cases. 
Of relevance for current and future missions is actually the integral of 
the distribution reported in Fig.~2. This, in fact, for each 
value reported on the x-axis, gives the probability that an 
observation will detect an X-ray flux due to the WHIM greater than that value. 
The result is shown in Fig. 3. From the figure we can see, for example that 
with a 3' FOV, in about 20\% of the observations we expect the WHIM flux to 
contribute to more than 20\% of the DXB (5.8 
$\textrm{~photons~cm}^{-2}\textrm{~s}^{-1}\textrm{~sr}^{-1}$).

This value alone is not sufficient to determine if a mission would be 
able to observe and study the X-ray flux from the WHIM. In fact, even 
when the X-ray flux is relatively high, if it's due to gas distributed at 
different redshifts it will show up in an observation almost as a continuum 
distribution, which is very difficult to separate from the other DXB 
flux. If, on the other end, the majority of the flux comes from gas at the 
same redshift (i.e., a single filament), then emission lines at that 
specific redshift are observable. We therefore split our FOV in small 
slices of redshift, approximately 0.03 thick (equivalent to the size of 
one Cen's cube), and we looked at the number of times that in each 
simulation the flux coming from a single slice was above a threshold 
equal to either 10\% or 20\% of the total DXB flux (i.e., 2.9 and 5.8 
$\textrm{~photons~cm}^{-2}\textrm{~s}^{-1}\textrm{~sr}^{-1}$). Figure 4 
shows the result of this analysis for the 3' FOV simulations. The result 
indicates that in almost 10\% of all the observations a single bright 
filament in the FOV accounts, alone, for more than 20\% of the DXB flux. 
This means that in almost 10\% of the cases we expect to have most of 
the WHIM flux concentrated at a single red-shift. We therefore 
expect to observe O~{\tiny VII} and O~{\tiny VIII} lines 
at that red-shift with an estimated flux of more than 5 
$\textrm{~photons~cm}^{-2}\textrm{~s}^{-1}\textrm{~sr}^{-1}$.
The same analysis repeated for 10' FOV and 30' FOV shows that, while the 
probability of having a high X-ray flux increases as the FOV is increased, 
the probability of having a significant amount of flux coming from a 
single redshift decreases drastically. To be able to detect and study the 
properties of the WHIM using X-ray flux an angular resolution of a few 
arcminutes is therefore necessary.

\section{Angular extension of the WHIM filaments}
The analysis of spectra from objects in a narrow field of view gives us 
useful information about the flux properties of the WHIM, and hints 
about its distribution, and detectability. However, to better understand 
the angular properties of the WHIM we also generated images in the soft 
X-ray band.

We chose a field of view of $30'\times30'$, adopting the same criteria we used for 
the spectral analysis, with a resolution of $32\times32$ pixels. The 
pixel size was chosen to reasonably match the dimension of the elements 
in Cen's cube. One single element of the box is 195 $h^{-1}$ kpc by side, 
equivalent to an angle of about 1 arcmin at a distance of 600 
$h^{-1}$ Mpc (redshift $\textrm{z}\sim0.18$). The basic idea is 
to create a three dimensional array. The first two axes represent the 
angular coordinates of the map, in arcminutes, and the third axis is the 
energy. Instead of a map we have a ``pile'' of maps, each representing 
the photon counts at increasing energy, from 0.05 to 2 keV. From this 
array is then easy to generate intensity or photon counts maps for different 
energy intervals. 

The procedure to generate the maps was essentially identical to the one 
discussed before, with each  pixel of the image corresponding to a 
separate FOV. Also in this case we saved one three-dimensional image for 
each of Cen's cubes 
in the field of view. This gives us pictures of the simulated universe in 
slices taken at increasing redshift. The simulated images can be used to 
extract information about the expected angular distribution of the WHIM 
and to understand the difficulties in extracting information about the 
WHIM properties from the analysis of real data. To quantify the angular 
distribution of the X-ray from the WHIM we generated images in the energy 
interval 0.375 - 0.950 keV and then we used an adaptation of the algorithm 
described by Soltan \cite{Soltan96} and updated by Kuntz \cite{Kuntz01} 
to calculate the Angular Autocorrelation Function (AcF) of the X-ray flux. 
The amount of 
clustering on different angular scales is quantified by the AcF:
\[ 
w(\theta)=\frac{\langle I(n)I(n')\rangle}{\langle I^2\rangle}, 
\]
where $I(n)$ is the intensity of the X-ray background (XRB) in the direction 
n, $I(n)I(n')$ is the product of intensities with angular separation 
$\theta$, and $\langle \ldots \rangle$ denotes the expectation values of the 
corresponding quantities. In X-ray astronomy, the radiation distribution 
is typically represented by an array of photon rates per pixel. As an 
estimator of the ACF, the expectation values are therefore replaced by 
the corresponding average quantities. A practical recipe to calculate 
the ACF, using a weighting of the intensity based on exposure time, 
is given by Kuntz (2001):
\[ 
w(\theta)=\frac{\sum_{i,i'}\Big(R-\overline{R}\Big)
 \Big(R'-\overline{R'}\Big)\sqrt{ss'}}{\sum_{i,i'}\sqrt{ss'}}\cdot
\frac{1}{\Big[\Big(\overline{R}+\overline{R'}\Big)/2\Big]^2},    
\] 
where the sum is over all pairs of pixels separated by $\theta$, $R$ is 
the photon rate in those pixels, s is the exposure time for those pixels, 
used as statistical weight, and $\overline{(\ldots)}$ denotes the average 
value. Notice that if a single image is used for the calculations, and 
$N_\theta$ is the number of pixels separated by $\theta$, then
$\overline{R}=\overline{R'}$, and we can write:
\begin{displaymath}
w(\theta)=\frac{\sum_{i=1,N_\theta}\sum_{j=i+1,N_\theta}
\Big(R_i-\overline{R}\Big)\Big(R_j-\overline{R} \Big)\sqrt{s_is_j}}
{\sum_{i=1,N_\theta}\sum_{j=i+1,N_\theta}\sqrt{s_is_j}}\cdot
\frac{1}{\overline{R}^2}.
\end{displaymath}	 

In Fig.~5 we show the autocorrelation function calculated for a sample 
of 40 images. The curves show a similar overall behavior, with some amount 
of dispersion that was expected from previous investigations \cite{Kuntz01}. 
Due to the extension of our images, the value of the AcF is reliable only 
between 1 arcmin up to about 20 arcmin, after which it degenerates due to 
the limited sampling of the images. Such interval seems to be more than 
appropriate to characterize the WHIM behavior. In fact, as the average of 
all the AcF shows (Fig.~6), there seem to be a clear angular correlation 
below a few arcminutes, while above 10 arcmin the curve is compatible with 
zero, indicating a typical angular scale of the WHIM filaments of a few 
arcminutes. This result and our previous considerations about the presence
of a single filament in the FOV indicate that any mission designed to 
investigate the WHIM should have an angular resolution not worse that a
few arcminutes. 

\section{Conclusions}

Based on the output of hydrodynamical simulations, we investigated the 
properties of the expected X-ray flux from the WHIM. This is of 
particular relevance in designing a mission optimized to detect and study 
the properties of the WHIM. Our simulations indicate that such a mission 
is possible and should be capable of detecting and characterize the WHIM. 
Moreover, in addition to good energy resolution (necessary to identify 
redshifted emission lines) and low background (due to the very low photon 
rate of the WHIM), our analysis shows that such a mission should have an 
angular resolution of at least a few arcminutes. 

\acknowledgments

We would like to thank Renyue Cen for giving public access to the output 
of his hydrodynamic simulation. We would also like to thank Mauro Roncarelli 
and Wilton T. Sanders for the useful discussion and suggestions.

\clearpage

\begin{figure}
\plotone{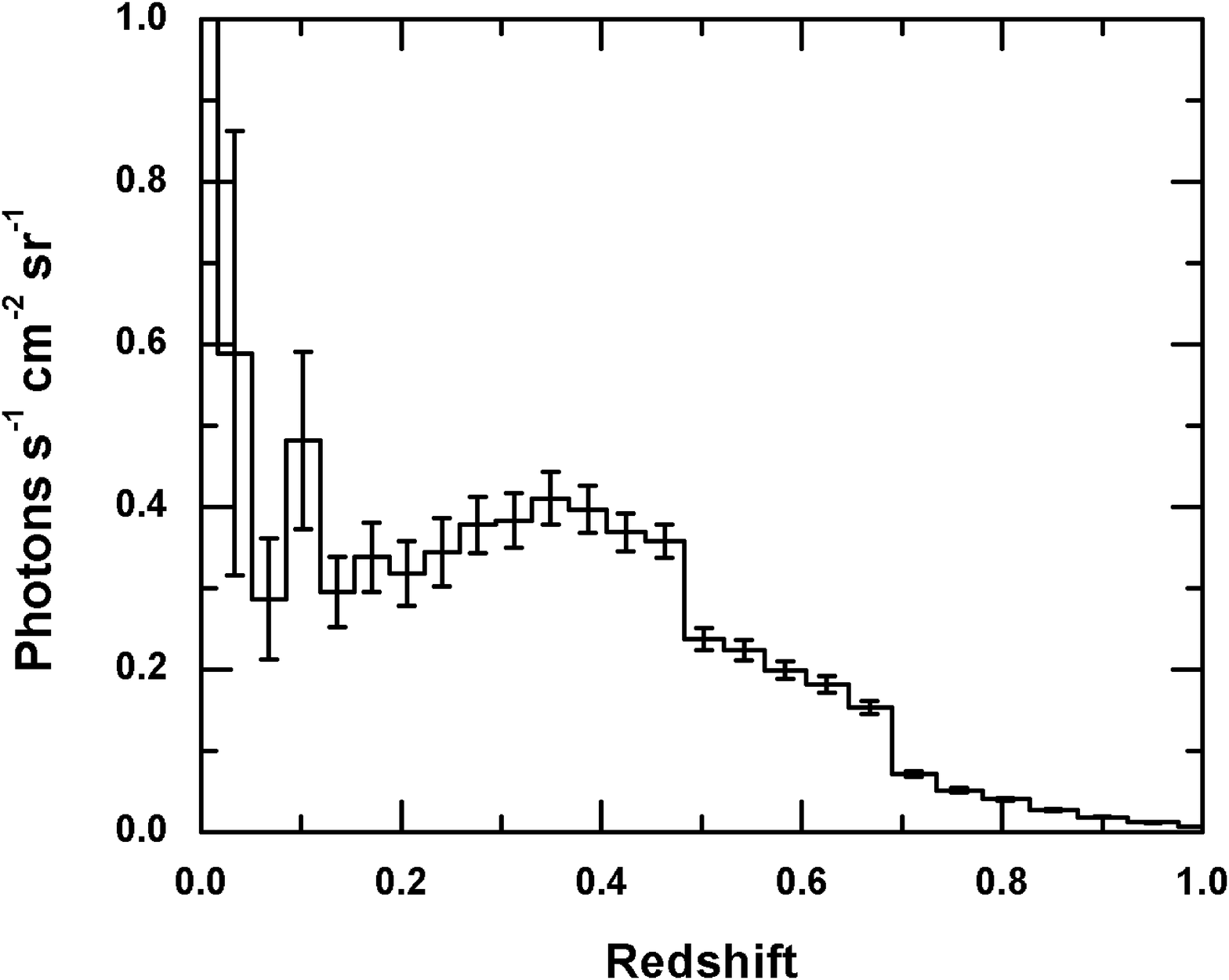}
\caption{Photon flux of the WHIM as a function of the distance from the 
emitter (redshift).\label{fig1}}
\end{figure}

\clearpage
\begin{figure}
\plotone{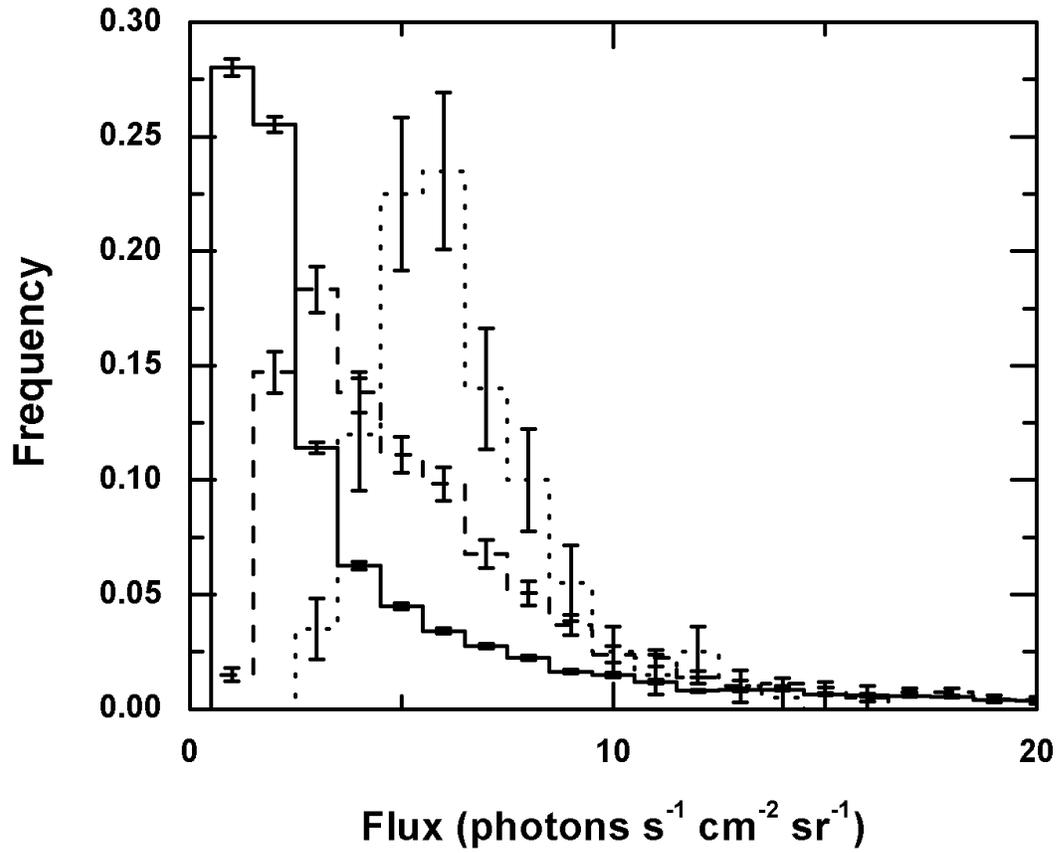}
\caption{Distribution of photon flux of the WHIM for simulations with FOV's of 
3'(solid line), 10' (dashed line), and 30' (dotted line). \label{fig2}}
\end{figure}

\clearpage
\begin{figure}
\plotone{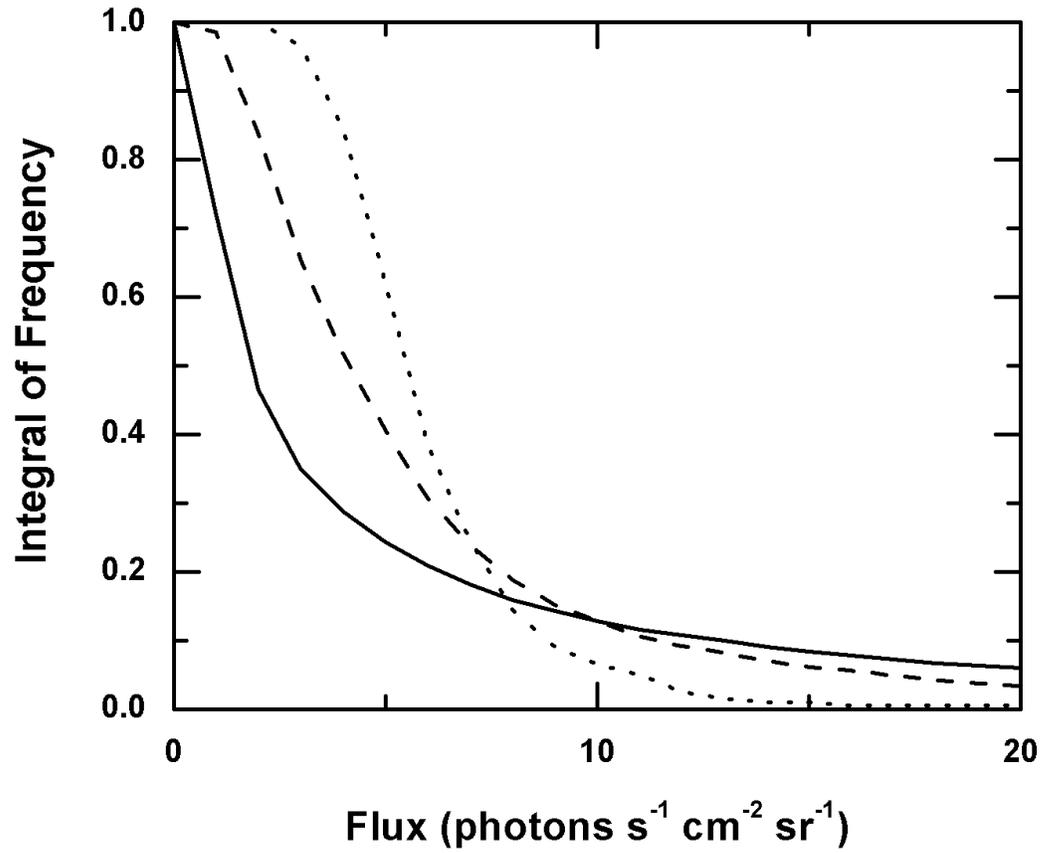}
\caption{Integral of the distribution of photon flux from Fig. 2 for simulations 
with FOV's of 3' (solid line), 10' (dashed line), and 30' 
(dotted line).\label{fig3}}
\end{figure}

\clearpage
\begin{figure}
\plotone{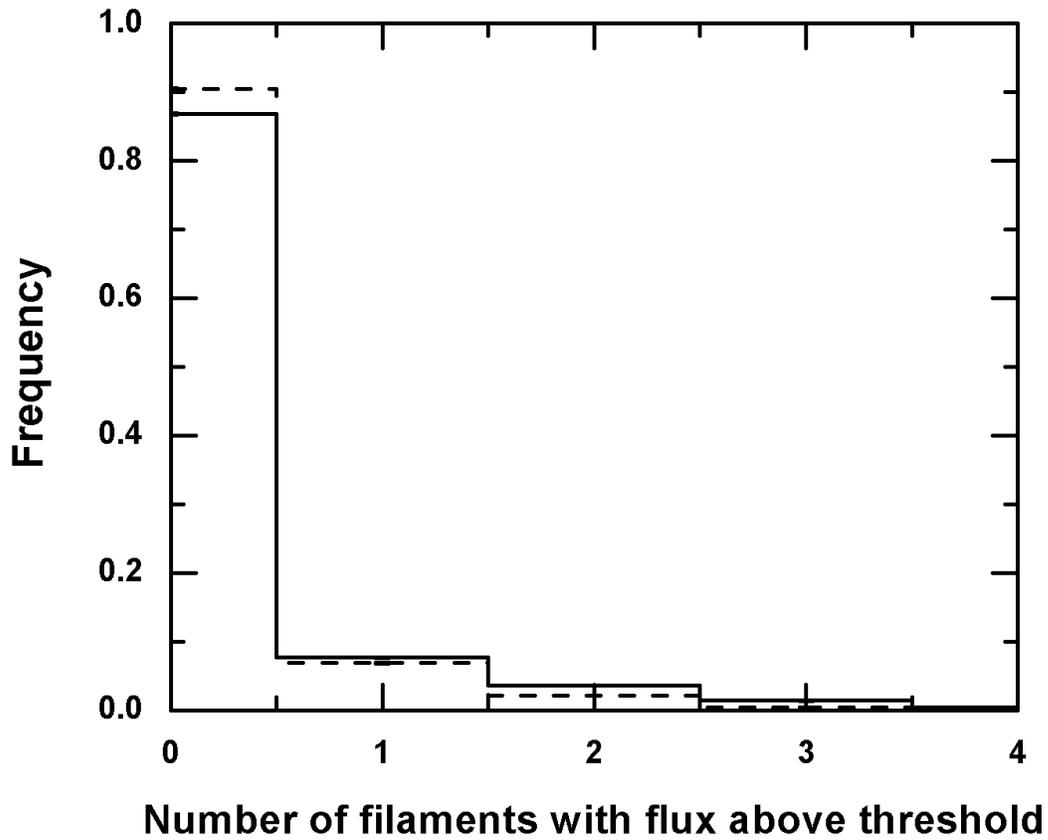}
\caption{Probability of observing, in one simulation, redshift intervals 
(filaments) with photon flux higher then 10\% (solid line) or 
20\% (dashed line) of the Diffuse X-Ray Background Flux. The data are for 
simulations with a FOV of 3'.\label{fig4}}
\end{figure}

\clearpage
\begin{figure}
\plotone{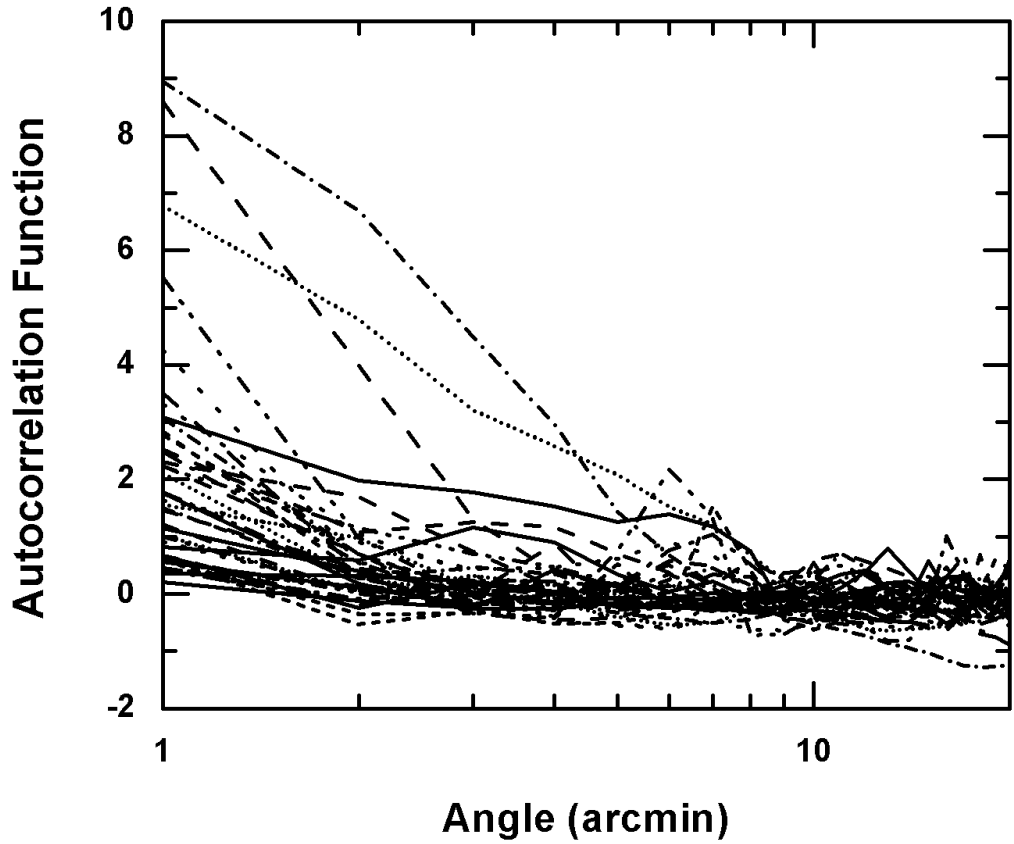}
\caption{Angular Autocorrelation functions of the X-ray in the energy interval
0.375 - 0.950 keV calculated for a sample of 40 images.
\label{fig5}}
\end{figure}

\clearpage
\begin{figure}
\plotone{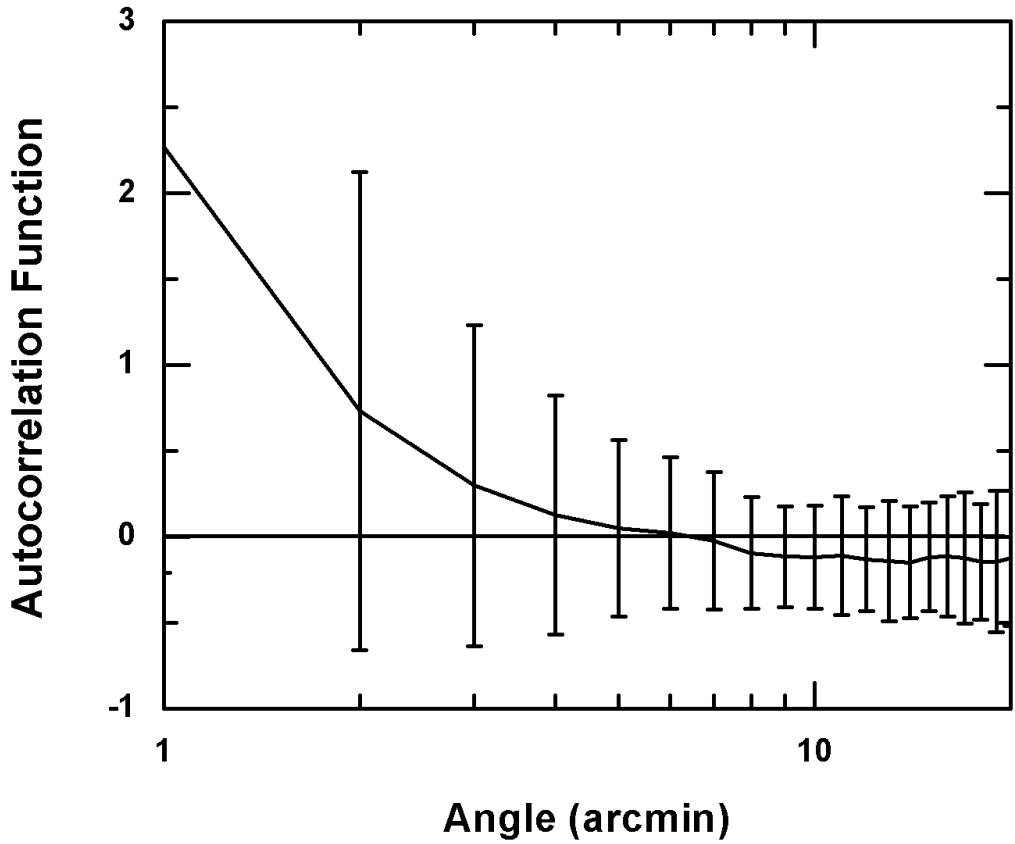}
\caption{Average angular autocorrelation function of the X-ray flux from the 
WHIM in the energy interval 0.375 - 0.950 keV.\label{fig6}}
\end{figure}

\clearpage

\end{document}